\documentclass[twocolumn,superscriptaddress,aps,showpacs,amsmath,footinbib,bibnotes]{revtex4-1}
\pdfoutput=1
\usepackage{amsmath}
\usepackage{amssymb}
\usepackage{stmaryrd}
\usepackage{mathrsfs}
\usepackage[utf8]{inputenc}
\usepackage[export]{adjustbox}
\usepackage[dvipsnames]{xcolor}

\begin{document}

\author{Amirhassan Shams-Ansari}
\affiliation{John A. Paulson School of Engineering and Applied Sciences, Harvard University, Cambridge, Massachusetts 02138, USA}
\affiliation{Department of Electrical Engineering and Computer Science, Howard University, Washington, DC 20059, USA}
\author{Pawel Latawiec}
\affiliation{John A. Paulson School of Engineering and Applied Sciences, Harvard University, Cambridge, Massachusetts 02138, USA}
\author{Yoshitomo Okawachi}
\affiliation{Department of Applied Physics and Applied Mathematics, Columbia University, New York, New York 10027, USA}
\author{Vivek Venkataraman}
\affiliation{John A. Paulson School of Engineering and Applied Sciences, Harvard University, Cambridge, Massachusetts 02138, USA}
\affiliation{Department of Electrical Engineering, Indian Institute of Technology, Delhi, New Delhi, India}
\author{Mengjie Yu}
\affiliation{Department of Applied Physics and Applied Mathematics, Columbia University, New York, New York 10027, USA}
\author{Boris Desiatov}
\affiliation{John A. Paulson School of Engineering and Applied Sciences, Harvard University, Cambridge, Massachusetts 02138, USA}
\author{Haig Atikian}
\affiliation{John A. Paulson School of Engineering and Applied Sciences, Harvard University, Cambridge, Massachusetts 02138, USA}
\author{Gary L. Harris}
\affiliation{Department of Electrical Engineering and Computer Science, Howard University, Washington, DC 20059, USA}
\author{Nathalie Picqu\'{e}}
\affiliation{Max-Planck Institute of Quantum Optics, Hans-Kopfermann-Str. 1, 85748 Garching, Germany, Germany}
\author{Alexander L. Gaeta}
\affiliation{Department of Applied Physics and Applied Mathematics, Columbia University, New York, New York 10027, USA}
\author{Marko Lon\v{c}ar}
\affiliation{John A. Paulson School of Engineering and Applied Sciences, Harvard University, Cambridge, Massachusetts 02138, USA}
\date{\today}
\title{Supercontinuum generation in angle-etched diamond waveguides}

\begin{abstract}
We experimentally demonstrate on-chip supercontinuum generation in the visible region in angle-etched diamond waveguides. We measure an output spectrum spanning 670 – 920 nm in a 5-mm-long waveguide using 100-fs pulses with 187 pJ of incident pulse energy. Our fabrication technique, combined with diamond’s broad transparency window, offers a potential route toward broadband supercontinuum generation in the UV domain.
\end{abstract}
\maketitle

Supercontinuum generation (SCG) is a nonlinear optical process, where a short, high-intensity pulse of light experiences significant spectral broadening due to combined effect of self- and cross-phase modulation, soliton fission, Cherenkov radiation, modulation instability, and Raman scattering \cite{agrawal2000nonlinear}. The development of supercontinuum (SC) sources has enabled numerous applications such as biomedical imaging, molecular detection, and high precision metrology \cite{dudley2006supercontinuum,strogatz2018nonlinear}. The generation of octave spanning SCG is of great importance, as it enables a fully stabilized frequency comb synthesizer through f-2f self-referencing \cite{diddams2000direct}. Recently, there has been significant development of on-chip SC sources in various platforms such as silicon \cite{leo2015coherent,leo2014dispersive,lau2014octave}, silica \cite{oh2014supercontinuum}, silicon-germanium \cite{sinobad2018mid}, aluminum nitride \cite{hickstein2017ultrabroadband,liu2019beyond}, silicon nitride \cite{lacava2017si,porcel2017two,johnson2015octave,klenner2016gigahertz,zhao2015visible,halir2012ultrabroadband}, chalcogenides \cite{lamont2008supercontinuum,xie20142}, periodically poled lithium niobate \cite{phillips2011supercontinuum}, and thin-film lithium niobate \cite{yu2019coherent,lu2019octave,Jankowski:19}, offering potential for compact devices that allow for low cost and large-scale fabrication.

There has been interest in extending the generated SC spectra into visible and ultra-violet (UV) wavelength range which can, for example, enhance the accuracy of optical coherence tomography \cite{fercher2003optical}. 
 However, observation of supercontinuum generation in UV and visible region in most materials are challenging due to the large normal material group-velocity dispersion (GVD). This dominates over efforts to engineer the dispersion via the waveguide geometry, making it challenging to achieve the anomalous GVD necessary for efficient phase-matched nonlinear optical processes. In addition, the intensified Rayleigh scattering at these wavelengths leads to challenges in making low-loss structures in the short wavelength regime.
  \begin{figure}
	\centering
	\includegraphics[width=\linewidth,height=\textheight,keepaspectratio]{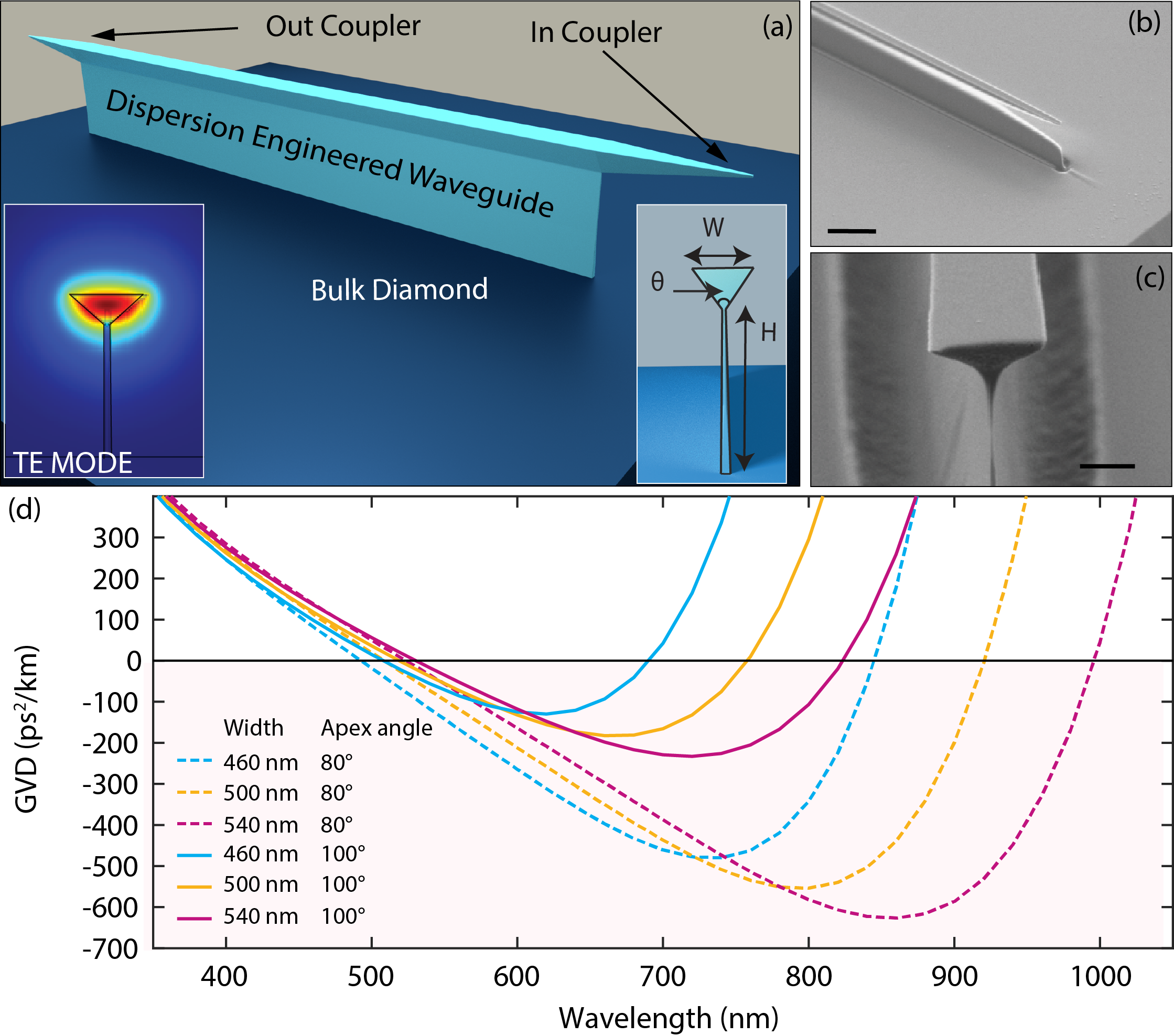}

	\caption{\label{fig1}\textbf{The angle-etched diamond waveguide. }\textbf{(a)} 3-D schematic of the device consisting of supported diamond waveguide with tapered sections at its ends, used ffor efficient in- and out coupling. Insets show the simulated TE mode in the case of W=540 nm, H=2 $\mu$m , $\theta=100^\circ$ (bottom left) and the waveguide cross-section with important parameters indicated: W is the width, H is the height, and $\theta$ is the apex angle of the waveguide (bottom right). \textbf{(b)} SEM image of the coupler (scale bar - 3 $\mu$m). \textbf{(c)} SEM image of the waveguide cross section (scale bar - 500 nm). \textbf{(d)} Simulated GVD for waveguides with apex angle of 100$^\circ$ (solid lines) and apex angle of 80$^\circ$ (dashed lines) and widths of 460 nm, 500 nm, and 540 nm. The region of anomalous GVD is shaded.}
\end{figure}

Diamond is an emerging platform for integrated photonics, owing to its wide bandgap which significantly reduces multi-photon absorption processes, its relatively high refractive index (\textit{n}=2.38), and its strong optical nonlinearities. Additionally, its structure can be engineered to have anomalous GVD in the telecom wavelength range \cite{feigel2017opportunities,hausmann2014diamond}. This has enabled realization of frequency combs \cite{hausmann2014diamond} and Raman lasers \cite{latawiec2015chip,latawiec2018integrated}, for example. Importantly, these demonstrations mainly relied on diamond on insulator (DOI) platform \cite{faraon2011resonant,hausmann2012integrated} that consists of a sub-micron thick single crystalline diamond (SCD) film, that is prepared using combination of polishing and reactive ion etching, transferred on top of a low index substrate such as fused silica or SiO2/Si. While this platform enabled important advances in the field of diamond photonics, it suffers from a low device yield and significant thickness variations across the chip with characteristic “wedge”-like profile introduced during polishing step. This is particularly problematic for nonlinear optical devices that rely on precise dispersion engineering. An alternative platform that we developed leverages angle-etching of diamond to realize free standing structures with sufficient optical isolation and has been used in quantum photonics and opto-mechanics \cite{meesala2016enhanced,burek2016diamond,sipahigil2016integrated,evans2018photon}. Since the structures are made directly from unpolished bulk diamond there is no thickness variation across the entire structure, making this technique advantageous over DOI. Angled-etched diamond platform has been the workhorse of diamond-based quantum photonics efforts \cite{meesala2016enhanced,sipahigil2016integrated,sohn2018controlling,sun2018cavity} but has not been utilized for realization of nonlinear optical devices. 

Here we leverage uniformity and scalability of reactive ion beam angled etching (RIBAE) \cite{atikian2017freestanding} approach to realize, for the first time, SCG in SCD. By controlling the geometry of diamond waveguide with characteristic triangular cross-section, and in particular its apex angle ($\theta$ in Figure 1(a) – right inset), we could achieve anomalous GVD over wide wavelength range (Figure 1(d)). Our numerical modeling indicates that our devices should support 250 nm wide SC spectrum spanning 670 – 920 nm (Figure 2(a) and (b)). 
 \begin{figure}[h!]
	\centering
	\includegraphics[width=\linewidth,height=\textheight,keepaspectratio]{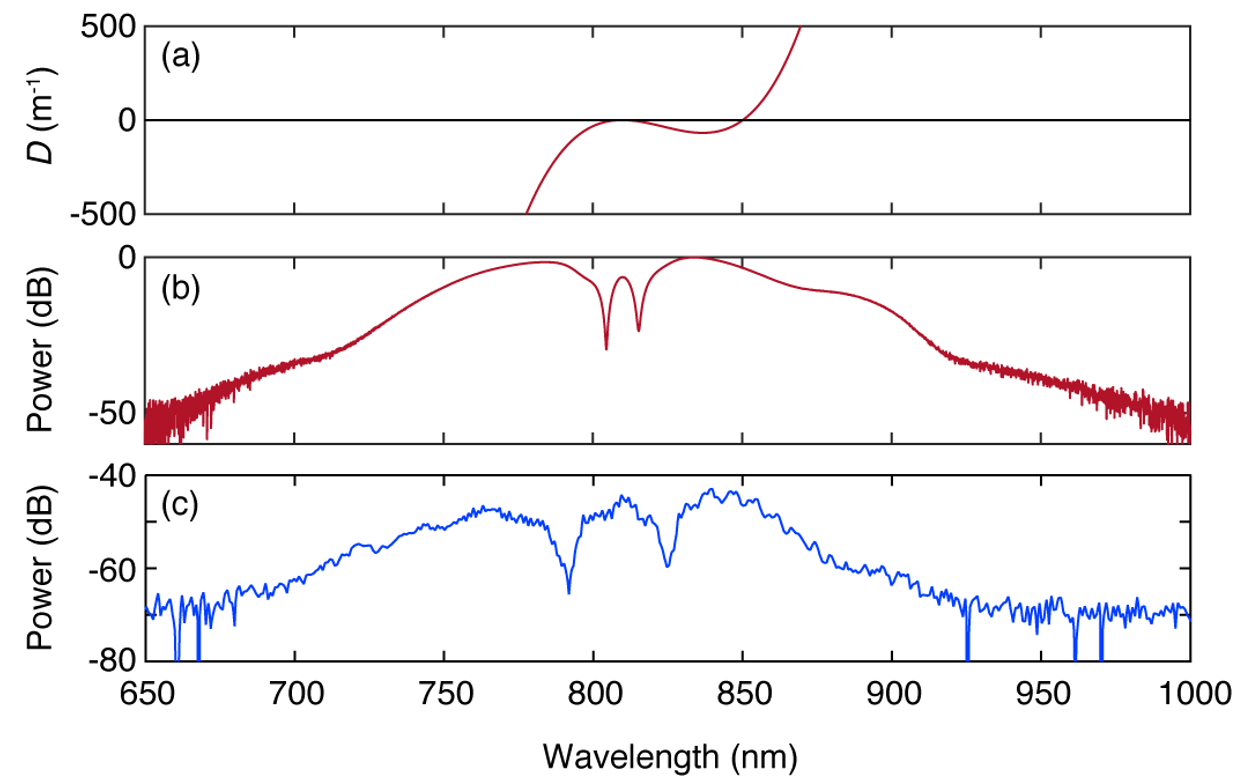}

	\caption{\label{fig1}\textbf{(a)} Dispersion operator for a waveguide with 540 nm width and 100$^\circ$ apex angle shown in Fig. 1.  \textbf{(b)} Simulated SC spectra for a 5 mm long waveguide with 540 nm width and 100$^\circ$ apex angle. \textbf{(c)} Experimental results for SCG in the waveguide for an input pulse energy of 187 pJ with 100 fs pulse centered at 810 nm and a repetition rate of 80 MHz.}
\end{figure}

The waveguide cross-section is that of an inverted triangle on a thin support pedestal [Figure 1(a)]. The pedestal provides mechanical support over the required waveguide length across the chip. The width of the support is 20 nm in order to prevent any mode leakage to the substrate. Figure 1(a) (left inset) shows the TE mode profile of a triangular diamond waveguide with an apex angle of 100 degree and width of 540 nm.  Dispersion was calculated for different waveguide widths and apex angles, using a finite element mode solver, and results are summarized in Figure 1 (d) in the case of the fundamental TE mode. By varying the widths and apex angle, the anomalous dispersion region can be tuned over a broad region.  Our design consists of two suspended tapered sections as in- and out-couplers surrounding the central fully supported dispersion-engineered region [Figure 1(a)- right inset]. The couplers (suspended regions) are designed to be 280 nm wide and 20 $\mu$m long to optimize the mode overlap with the input field. 

Devices are fabricated using mechanically polished $\sim$ 5$\times$5 mm – 500 $\mu$m thick electronic-grade SCD with a [001]-oriented surface (Element Six). After cleaning in a boiling mixture of HClO\textsubscript{4}:H\textsubscript{2}SO\textsubscript{4}:HNO\textsubscript{3} (1:1:1) followed by solvent cleaning, a layer of Nb is deposited as hard-mask using magnetron sputtering (AJA-ATC). Waveguides of a length of 5mm are written using E-beam lithography with multi-pass exposure (Elionix –F125) with negative-tone e-beam resist (FOx-16, Dow Corning). This pattern is then transferred first to Nb using an Ar/Cl\textsubscript{2} etch chemistry, and then to diamond using an inductively coupled plasma-reactive ion etcher (ICP-RIE, Unaxis Plasma-Therm) with O\textsubscript{2}. This results in 5 mm long and 2 $\mu$m thick diamond ridges. After this top-down etch, the sample is placed in a reactive ion-beam etcher and stage tilt was selected to result in the desired apex angle. Oxygen ion beam was used to perform the etch. Finally, the Nb mask was removed in H\textsubscript{3}PO\textsubscript{4}:H\textsubscript{2}SO\textsubscript{4}: HNO\textsubscript{3} (1:1:1) and device is cleaned in H\textsubscript{2}SO\textsubscript{4}:H\textsubscript{2}O\textsubscript{2} (3:1) followed by solvent clean and dried in a critical point drier to prevent damaging the suspended in-out couplers. Because the diamond cannot be cleaved or polished after the fabrication is complete, extra steps are performed to ensure the waveguides are written as close to the edge as possible (5-10 $\mu$m) to allow for efficient in-coupling of light using a lens. These steps consist of oxygen-plasma surface treatments and the use of adhesion promoters before spinning. Scanning electron micrographs (SEM) of the coupling section and cross section of a fabricated waveguide are shown in the inset of Figure1(b) and Figure 1(c), respectively.

We model the pulse propagation dynamics in the diamond waveguide using the split-step Fourier technique to solve the generalized nonlinear envelope equation. We consider the contributions from higher-order dispersion, third-order nonlinearity, and self-steepening in the model. The dispersion operator, $D=\sum_{n=2,3,\ldots}\frac{\beta_n(\omega_0)}{n!}(\omega-\omega_0)^n$ shown in figure 2(a), predicts the onset of a dispersive wave near 850 nm \cite{okawachi2017coherent}. Figure 2(b) shows the simulated spectrum for a 5-mm-length diamond waveguide with a 540 nm width and a 100$^\circ$ apex angle. In our simulation we assumed a 100-fs pump with a pulse energy of 21 pJ in the waveguide. However, we observe that the spectral broadening is largely due to self-phase modulation \cite{agrawal2000nonlinear}.  

In our experiment, the angle-etched waveguides are pumped with a Ti: Sapphire laser centered at 810 nm with a pulse duration of 100 fs and a repetition rate of 80 MHz. A variable neutral density filter is used to control the pump power. We use an aspheric lens for coupling into the waveguide and collect the output using a lensed fiber and send it to an optical spectrum analyzer (OSA). Figure 2(c) shows the generated SC spectrum. For 187 pJ of incident pulse energy, we observe a broadband spectrum spanning 250 nm from 670– 920 nm. Our measured spectrum shows good agreement with the simulated spectrum. The discrepancy in spectral position is attributed to fabrication tolerances in the waveguide width.
 \begin{figure}
	\centering
	\includegraphics[width=\linewidth,height=\textheight,keepaspectratio]{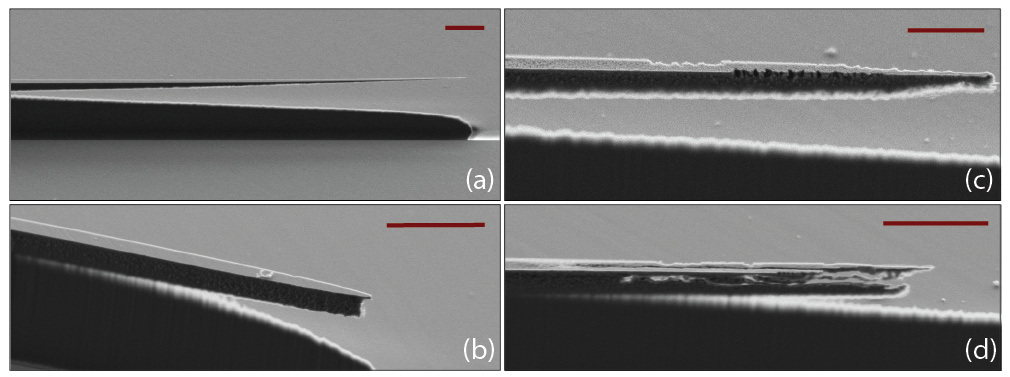}

	\caption{\label{fig1}\textbf{(a,b)} Coupling region of the waveguide before damage \textbf{(c,d)} damaged waveguide after exposure to high peak energy  - scale bars represent 2$\mu$m.}
\end{figure}
The low transmission of the device can be attributed to imperfections in the coupling facet and the sidewall roughness of the waveguides. By comparing the simulation and the experiment, a coupling loss of ~10 dB/ facet is calculated. In addition, we observe damage to the waveguides facet for high pump powers. We attribute this to excessive peak- intensity leading to femto-second ablation of diamond which can happen with photons with energies below diamonds bandgap in the presence of ultra-short pulses \cite{dumitru2002femtosecond,wu2003femtosecond,shinoda2009femtosecond}. This results in graphitization phenomenon and incubation effects starting from the coupling region and extending toward the rest of the waveguide [Figure 3].

As it was discussed before, diamond allows for dispersion engineering in the visible and UV region. To confirm this, we numerically design angle-etched structures which would extend the SC spectra toward the UV range. To shift the region of anomalous GVD to even lower wavelengths, we consider the 80$^\circ$ apex angle with narrower widths. Figure 4(a) shows the GVD profile for an angle-etched waveguide with apex angle of 50$^\circ$ and a 250 nm width and 4(b) shows the corresponding dispersion operator for a pump wavelength of 520 nm. The dispersion operator indicates that two dispersive waves can be generated, one near 260 nm and the other near 650 nm. Figure 4(c) shows the simulated SC spectrum. We neglect the Raman effect for TE polarization assuming propagation along the (100) axis \cite{mildren2013optical}. Assisted by the dispersive wave, the SC spectrum covers over an octave of bandwidth, well into the UV. However, in our experiment we were not able to see any broadening because of the fast degradation of the waveguides exposed to 520nm laser light. We project that optimization of the waveguide design and length, as well as coupler design, will allow for lower required pulse energies, preventing taper damage.
 \begin{figure}[h!]
	\centering
	\includegraphics[width=\linewidth,height=\textheight,keepaspectratio]{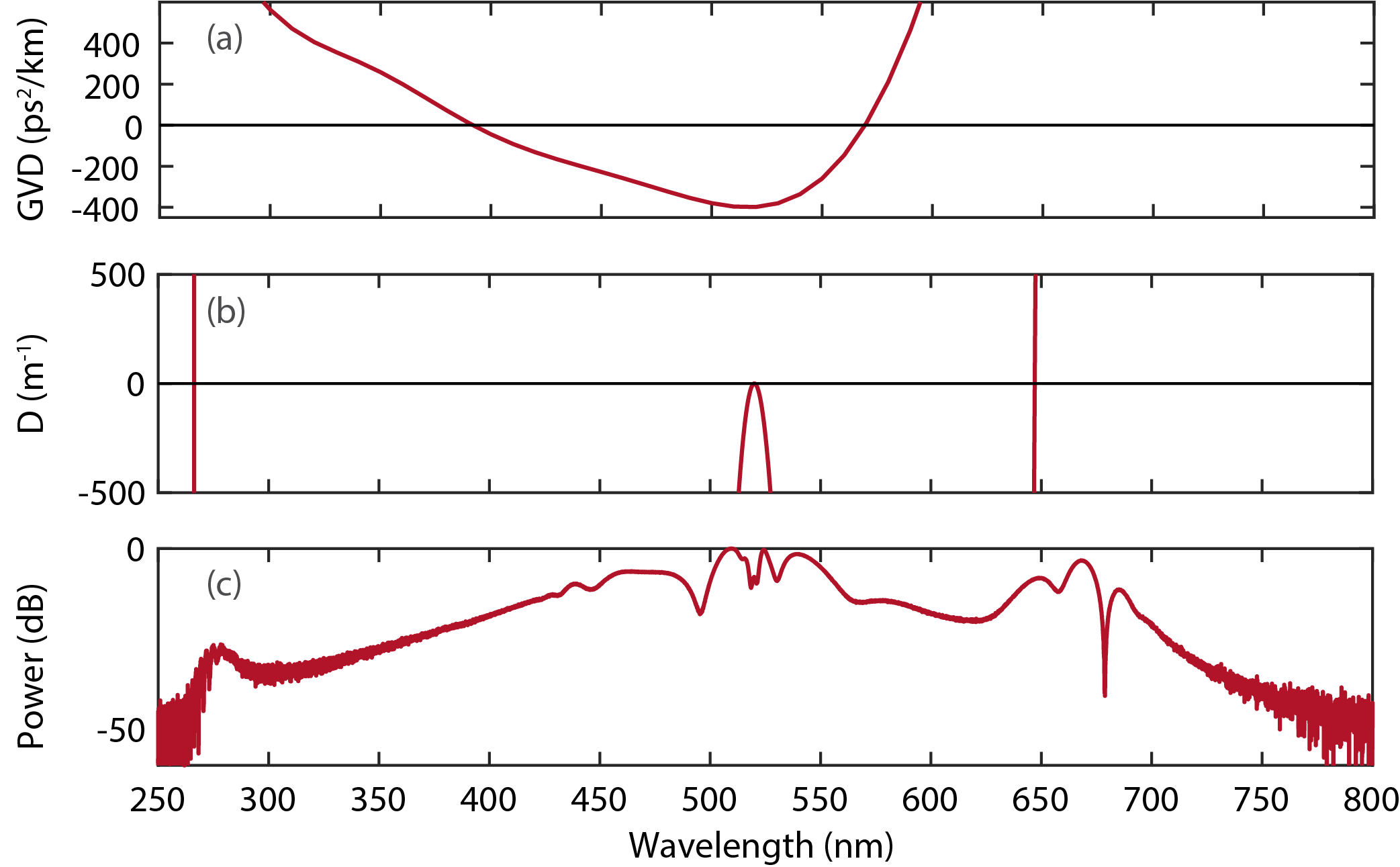}

	\caption{\label{fig1}\textbf{(a)} Simulated GVD for an angle-etched waveguide with apex angle of 50$^\circ$ and a 250 nm width. \textbf{(b)} Dispersion operator for a 520 nm pump wavelength.\textbf{(c)} Simulated supercontinuum spectrum for 100 fs pulse with 31.2 pJ of pulse energy.}
\end{figure}

In conclusion, by leveraging fabrication advances in angled-etching toward a novel integrated optics device architecture, we have demonstrated the first SCG in SC-diamond in the visible region spanning 670-920 nm. Owing to diamond’s material properties, the proposed design has the potential of generating a SC spectrum extending into the UV. Even though, achieving a broad spectrum in diamond is challenging, but our angle-etched design provides a new way for dispersion engineering that can be used for other integrated platforms.

\bibliographystyle{ieeetr}
\bibliography{bibliography}

\begin{thebibliography}{10}

\bibitem{agrawal2000nonlinear}
G.~P. Agrawal, ``Nonlinear fiber optics,'' in {\em Nonlinear Science at the
  Dawn of the 21st Century}, pp.~195--211, Springer, 2000.

\bibitem{dudley2006supercontinuum}
J.~M. Dudley, G.~Genty, and S.~Coen, ``Supercontinuum generation in photonic
  crystal fiber,'' {\em Reviews of modern physics}, vol.~78, no.~4, p.~1135,
  2006.

\bibitem{strogatz2018nonlinear}
S.~H. Strogatz, {\em Nonlinear Dynamics and Chaos with Student Solutions
  Manual: With Applications to Physics, Biology, Chemistry, and Engineering}.
\newblock CRC Press, 2018.

\bibitem{diddams2000direct}
S.~A. Diddams, D.~J. Jones, J.~Ye, S.~T. Cundiff, J.~L. Hall, J.~K. Ranka,
  R.~S. Windeler, R.~Holzwarth, T.~Udem, and T.~H{\"a}nsch, ``Direct link
  between microwave and optical frequencies with a 300 thz femtosecond laser
  comb,'' {\em Physical Review Letters}, vol.~84, no.~22, p.~5102, 2000.

\bibitem{leo2015coherent}
F.~Leo, S.-P. Gorza, S.~Coen, B.~Kuyken, and G.~Roelkens, ``Coherent
  supercontinuum generation in a silicon photonic wire in the telecommunication
  wavelength range,'' {\em Optics letters}, vol.~40, no.~1, pp.~123--126, 2015.

\bibitem{leo2014dispersive}
F.~Leo, S.-P. Gorza, J.~Safioui, P.~Kockaert, S.~Coen, U.~Dave, B.~Kuyken, and
  G.~Roelkens, ``Dispersive wave emission and supercontinuum generation in a
  silicon wire waveguide pumped around the 1550 nm telecommunication
  wavelength,'' {\em Optics letters}, vol.~39, no.~12, pp.~3623--3626, 2014.

\bibitem{lau2014octave}
R.~K. Lau, M.~R. Lamont, A.~G. Griffith, Y.~Okawachi, M.~Lipson, and A.~L.
  Gaeta, ``Octave-spanning mid-infrared supercontinuum generation in silicon
  nanowaveguides,'' {\em Optics letters}, vol.~39, no.~15, pp.~4518--4521,
  2014.

\bibitem{oh2014supercontinuum}
D.~Y. Oh, D.~Sell, H.~Lee, K.~Y. Yang, S.~A. Diddams, and K.~J. Vahala,
  ``Supercontinuum generation in an on-chip silica waveguide,'' {\em Optics
  letters}, vol.~39, no.~4, pp.~1046--1048, 2014.

\bibitem{sinobad2018mid}
M.~Sinobad, C.~Monat, B.~Luther-Davies, P.~Ma, S.~Madden, D.~J. Moss,
  A.~Mitchell, D.~Allioux, R.~Orobtchouk, S.~Boutami, {\em et~al.},
  ``Mid-infrared octave spanning supercontinuum generation to 8.5 $\mu$m in
  silicon-germanium waveguides,'' {\em Optica}, vol.~5, no.~4, pp.~360--366,
  2018.

\bibitem{hickstein2017ultrabroadband}
D.~D. Hickstein, H.~Jung, D.~R. Carlson, A.~Lind, I.~Coddington, K.~Srinivasan,
  G.~G. Ycas, D.~C. Cole, A.~Kowligy, C.~Fredrick, {\em et~al.},
  ``Ultrabroadband supercontinuum generation and frequency-comb stabilization
  using on-chip waveguides with both cubic and quadratic nonlinearities,'' {\em
  Physical Review Applied}, vol.~8, no.~1, p.~014025, 2017.

\bibitem{liu2019beyond}
X.~Liu, A.~W. Bruch, J.~Lu, Z.~Gong, J.~B. Surya, L.~Zhang, J.~Wang, J.~Yan,
  and H.~X. Tang, ``Beyond 100 thz-spanning ultraviolet frequency combs in a
  non-centrosymmetric crystalline waveguide,'' {\em arXiv preprint
  arXiv:1906.00323}, 2019.

\bibitem{lacava2017si}
C.~Lacava, S.~Stankovic, A.~Z. Khokhar, T.~D. Bucio, F.~Gardes, G.~T. Reed,
  D.~J. Richardson, and P.~Petropoulos, ``Si-rich silicon nitride for nonlinear
  signal processing applications,'' {\em Scientific reports}, vol.~7, no.~1,
  p.~22, 2017.

\bibitem{porcel2017two}
M.~A. Porcel, F.~Schepers, J.~P. Epping, T.~Hellwig, M.~Hoekman, R.~G.
  Heideman, P.~J. van~der Slot, C.~J. Lee, R.~Schmidt, R.~Bratschitsch, {\em
  et~al.}, ``Two-octave spanning supercontinuum generation in stoichiometric
  silicon nitride waveguides pumped at telecom wavelengths,'' {\em Optics
  express}, vol.~25, no.~2, pp.~1542--1554, 2017.

\bibitem{johnson2015octave}
A.~R. Johnson, A.~S. Mayer, A.~Klenner, K.~Luke, E.~S. Lamb, M.~R. Lamont,
  C.~Joshi, Y.~Okawachi, F.~W. Wise, M.~Lipson, {\em et~al.}, ``Octave-spanning
  coherent supercontinuum generation in a silicon nitride waveguide,'' {\em
  Optics letters}, vol.~40, no.~21, pp.~5117--5120, 2015.

\bibitem{klenner2016gigahertz}
A.~Klenner, A.~S. Mayer, A.~R. Johnson, K.~Luke, M.~R. Lamont, Y.~Okawachi,
  M.~Lipson, A.~L. Gaeta, and U.~Keller, ``Gigahertz frequency comb offset
  stabilization based on supercontinuum generation in silicon nitride
  waveguides,'' {\em Optics express}, vol.~24, no.~10, pp.~11043--11053, 2016.

\bibitem{zhao2015visible}
H.~Zhao, B.~Kuyken, S.~Clemmen, F.~Leo, A.~Subramanian, A.~Dhakal, P.~Helin,
  S.~Severi, E.~Brainis, G.~Roelkens, {\em et~al.}, ``Visible-to-near-infrared
  octave spanning supercontinuum generation in a silicon nitride waveguide,''
  {\em Optics letters}, vol.~40, no.~10, pp.~2177--2180, 2015.

\bibitem{halir2012ultrabroadband}
R.~Halir, Y.~Okawachi, J.~Levy, M.~Foster, M.~Lipson, and A.~Gaeta,
  ``Ultrabroadband supercontinuum generation in a cmos-compatible platform,''
  {\em Optics letters}, vol.~37, no.~10, pp.~1685--1687, 2012.

\bibitem{lamont2008supercontinuum}
M.~R. Lamont, B.~Luther-Davies, D.-Y. Choi, S.~Madden, and B.~J. Eggleton,
  ``Supercontinuum generation in dispersion engineered highly nonlinear
  ($\gamma$= 10/w/m) as 2 s 3 chalcogenide planar waveguide,'' {\em Optics
  Express}, vol.~16, no.~19, pp.~14938--14944, 2008.

\bibitem{xie20142}
S.~Xie, F.~Tani, J.~C. Travers, P.~Uebel, C.~Caillaud, J.~Troles, M.~A.
  Schmidt, and P.~S.~J. Russell, ``As 2 s 3--silica double-nanospike waveguide
  for mid-infrared supercontinuum generation,'' {\em Optics letters}, vol.~39,
  no.~17, pp.~5216--5219, 2014.

\bibitem{phillips2011supercontinuum}
C.~Phillips, C.~Langrock, J.~Pelc, M.~Fejer, J.~Jiang, M.~E. Fermann, and
  I.~Hartl, ``Supercontinuum generation in quasi-phase-matched linbo 3
  waveguide pumped by a tm-doped fiber laser system,'' {\em Optics letters},
  vol.~36, no.~19, pp.~3912--3914, 2011.

\bibitem{yu2019coherent}
M.~Yu, B.~Desiatov, Y.~Okawachi, A.~L. Gaeta, and M.~Lon{\v{c}}ar, ``Coherent
  two-octave-spanning supercontinuum generation in lithium-niobate
  waveguides,'' {\em Optics letters}, vol.~44, no.~5, pp.~1222--1225, 2019.

\bibitem{lu2019octave}
J.~Lu, J.~B. Surya, X.~Liu, Y.~Xu, and H.~X. Tang, ``Octave-spanning
  supercontinuum generation in nanoscale lithium niobate waveguides,'' {\em
  Optics letters}, vol.~44, no.~6, pp.~1492--1495, 2019.

\bibitem{Jankowski:19}
M.~Jankowski, C.~Langrock, B.~Desiatov, A.~Marandi, C.~Wang, M.~Zhang, C.~R.
  Phillips, M.~Loncar, and M.~M. Fejer, ``Ultrabroadband nonlinear optics in
  dispersion engineered periodically poled lithium niobate waveguides,'' in
  {\em Conference on Lasers and Electro-Optics}, p.~SM3O.2, Optical Society of
  America, 2019.

\bibitem{fercher2003optical}
A.~F. Fercher, W.~Drexler, C.~K. Hitzenberger, and T.~Lasser, ``Optical
  coherence tomography-principles and applications,'' {\em Reports on progress
  in physics}, vol.~66, no.~2, p.~239, 2003.

\bibitem{feigel2017opportunities}
B.~Feigel, D.~Castell{\'o}-Lurbe, H.~Thienpont, and N.~Vermeulen,
  ``Opportunities for visible supercontinuum light generation in integrated
  diamond waveguides,'' {\em Optics letters}, vol.~42, no.~19, pp.~3804--3807,
  2017.

\bibitem{hausmann2014diamond}
B.~Hausmann, I.~Bulu, V.~Venkataraman, P.~Deotare, and M.~Lon{\v{c}}ar,
  ``Diamond nonlinear photonics,'' {\em Nature Photonics}, vol.~8, no.~5,
  p.~369, 2014.

\bibitem{latawiec2015chip}
P.~Latawiec, V.~Venkataraman, M.~J. Burek, B.~J. Hausmann, I.~Bulu, and
  M.~Lon{\v{c}}ar, ``On-chip diamond raman laser,'' {\em Optica}, vol.~2,
  no.~11, pp.~924--928, 2015.

\bibitem{latawiec2018integrated}
P.~Latawiec, V.~Venkataraman, A.~Shams-Ansari, M.~Markham, and M.~Lon{\v{c}}ar,
  ``Integrated diamond raman laser pumped in the near-visible,'' {\em Optics
  letters}, vol.~43, no.~2, pp.~318--321, 2018.

\bibitem{faraon2011resonant}
A.~Faraon, P.~E. Barclay, C.~Santori, K.-M.~C. Fu, and R.~G. Beausoleil,
  ``Resonant enhancement of the zero-phonon emission from a colour centre in a
  diamond cavity,'' {\em Nature Photonics}, vol.~5, no.~5, p.~301, 2011.

\bibitem{hausmann2012integrated}
B.~J. Hausmann, B.~Shields, Q.~Quan, P.~Maletinsky, M.~McCutcheon, J.~T. Choy,
  T.~M. Babinec, A.~Kubanek, A.~Yacoby, M.~D. Lukin, {\em et~al.}, ``Integrated
  diamond networks for quantum nanophotonics,'' {\em Nano letters}, vol.~12,
  no.~3, pp.~1578--1582, 2012.

\bibitem{meesala2016enhanced}
S.~Meesala, Y.-I. Sohn, H.~A. Atikian, S.~Kim, M.~J. Burek, J.~T. Choy, and
  M.~Lon{\v{c}}ar, ``Enhanced strain coupling of nitrogen-vacancy spins to
  nanoscale diamond cantilevers,'' {\em Physical Review Applied}, vol.~5,
  no.~3, p.~034010, 2016.

\bibitem{burek2016diamond}
M.~J. Burek, J.~D. Cohen, S.~M. Meenehan, N.~El-Sawah, C.~Chia, T.~Ruelle,
  S.~Meesala, J.~Rochman, H.~A. Atikian, M.~Markham, {\em et~al.}, ``Diamond
  optomechanical crystals,'' {\em Optica}, vol.~3, no.~12, pp.~1404--1411,
  2016.

\bibitem{sipahigil2016integrated}
A.~Sipahigil, R.~E. Evans, D.~D. Sukachev, M.~J. Burek, J.~Borregaard, M.~K.
  Bhaskar, C.~T. Nguyen, J.~L. Pacheco, H.~A. Atikian, C.~Meuwly, {\em et~al.},
  ``An integrated diamond nanophotonics platform for quantum-optical
  networks,'' {\em Science}, vol.~354, no.~6314, pp.~847--850, 2016.

\bibitem{evans2018photon}
R.~E. Evans, M.~K. Bhaskar, D.~D. Sukachev, C.~T. Nguyen, A.~Sipahigil, M.~J.
  Burek, B.~Machielse, G.~H. Zhang, A.~S. Zibrov, E.~Bielejec, {\em et~al.},
  ``Photon-mediated interactions between quantum emitters in a diamond
  nanocavity,'' {\em Science}, vol.~362, no.~6415, pp.~662--665, 2018.

\bibitem{sohn2018controlling}
Y.-I. Sohn, S.~Meesala, B.~Pingault, H.~A. Atikian, J.~Holzgrafe,
  M.~G{\"u}ndo{\u{g}}an, C.~Stavrakas, M.~J. Stanley, A.~Sipahigil, J.~Choi,
  {\em et~al.}, ``Controlling the coherence of a diamond spin qubit through its
  strain environment,'' {\em Nature communications}, vol.~9, no.~1, p.~2012,
  2018.

\bibitem{sun2018cavity}
S.~Sun, J.~L. Zhang, K.~A. Fischer, M.~J. Burek, C.~Dory, K.~G. Lagoudakis,
  Y.-K. Tzeng, M.~Radulaski, Y.~Kelaita, A.~Safavi-Naeini, {\em et~al.},
  ``Cavity-enhanced raman emission from a single color center in a solid,''
  {\em Physical review letters}, vol.~121, no.~8, p.~083601, 2018.

\bibitem{atikian2017freestanding}
H.~A. Atikian, P.~Latawiec, M.~J. Burek, Y.-I. Sohn, S.~Meesala, N.~Gravel,
  A.~B. Kouki, and M.~Lon{\v{c}}ar, ``Freestanding nanostructures via reactive
  ion beam angled etching,'' {\em APL Photonics}, vol.~2, no.~5, p.~051301,
  2017.

\bibitem{okawachi2017coherent}
Y.~Okawachi, M.~Yu, J.~Cardenas, X.~Ji, M.~Lipson, and A.~L. Gaeta, ``Coherent,
  directional supercontinuum generation,'' {\em Optics letters}, vol.~42,
  no.~21, pp.~4466--4469, 2017.

\bibitem{dumitru2002femtosecond}
G.~Dumitru, V.~Romano, H.~Weber, M.~Sentis, and W.~Marine, ``Femtosecond
  ablation of ultrahard materials,'' {\em Applied Physics A}, vol.~74, no.~6,
  pp.~729--739, 2002.

\bibitem{wu2003femtosecond}
Q.~Wu, Y.~Ma, R.~Fang, Y.~Liao, Q.~Yu, X.~Chen, and K.~Wang, ``Femtosecond
  laser-induced periodic surface structure on diamond film,'' {\em Applied
  Physics Letters}, vol.~82, no.~11, pp.~1703--1705, 2003.

\bibitem{shinoda2009femtosecond}
M.~Shinoda, R.~R. Gattass, and E.~Mazur, ``Femtosecond laser-induced formation
  of nanometer-width grooves on synthetic single-crystal diamond surfaces,''
  {\em Journal of Applied Physics}, vol.~105, no.~5, p.~053102, 2009.

\bibitem{mildren2013optical}
R.~Mildren and J.~Rabeau, {\em Optical engineering of diamond}.
\newblock John Wiley \& Sons, 2013.

\end{thebibliography}

\end{document}